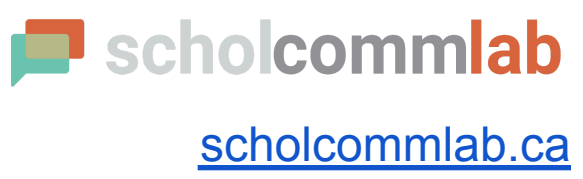

scholcommlab.ca

# Diamond Fractures: Tracing Journal Transitions Away from Diamond Open Access

Lisa Matthias[1,2,*], Juan Pablo Alperin[2,3], Mikael Laakso[4]

[1] Humboldt-Universität zu Berlin, Berlin (Germany)

[2] Scholarly Communications Lab, Ottawa/Vancouver (Canada)

[3] Simon Fraser University, Vancouver (Canada)

[4] Tampere University, Tampere (Finland)

*Corresponding author: l.a.matthia@gmail.com

l.a.matthia@gmail.com - https://orcid.org/0000-0002-2612-2132

juan@alperin.ca - https://orcid.org/0000-0002-9344-7439

mikael.laakso@tuni.fi - https://orcid.org/0000-0003-3951-7990

## Abstract

While much attention has been paid to journals transitioning toward Diamond Open Access (OA), comparatively little is known about those moving in the opposite direction. We introduce the concept of “diamond fractures”—instances in which journals that once published freely for both readers and authors subsequently abandoned that model, transitioning to subscription-based or article processing charge (APC)-funded publishing. Drawing on publisher OA portfolio records, historical APC lists, removal logs from the Directory of Open Access Journals (DOAJ), and a survey of journals using Open Journal Systems (OJS), we identified and characterized more than 440 journals that have undergone such fractures and analyzed how they are distributed across publisher types, disciplines, and regions; which models journals transition to and at what price points; how old journals are when they fracture; and how publication volume shifts in the period surrounding the transition. This paper reports on more than 440 confirmed fractures, the majority of which involve transitions to charging APCs, ranging from $8 to $5,300. These findings raise urgent questions about the stability of diamond OA as a publishing ecosystem. The transition to APCs—even at initially modest price points—follows well-documented warnings about the hyperinflationary tendencies of author-pays models, in which charges introduced as a pragmatic stopgap have repeatedly escalated over time. The rate of fractures are likely to intensify as APC-based publishing continues to be normalized through funder mandates and commercial OA incentives. Diamond fractures demand to be treated not as isolated institutional decisions but as a systemic

risk—one that can only be addressed through sustained investment in community-governed infrastructure and funding models that reduce the conditions under which abandoning diamond OA becomes the path of least resistance.

## Introduction

The early open access (OA) movement was driven primarily by a question of access: How to make research freely available to readers. Under this logic, article processing charges (APCs) offered a practical mechanism—shifting the cost of publishing from readers to authors and their institutions, thereby removing paywalls without disrupting existing publishing infrastructure. Despite early warnings, particularly from scholars in Latin America and the Global South who argued that commercializing scholarly communications risked replicating the barriers the OA movement sought to dismantle (CLASCO, 2015), APCs proved commercially attractive and spread rapidly. APC-based publishing has since become a dominant model and the primary mechanism through which funders and institutions operationalize their OA mandates, resulting in billions in annual revenue for publishers (Butler et al., 2023; Haustein et al., 2024).

In recent years, pushback against the APC model has reached a tipping point, galvanizing efforts to promote alternatives. Diamond OA, in which journals charge neither readers nor authors, has emerged as a central focus. This momentum is reflected in multiple diamond OA Summits, where proponents of this model have gathered to discuss how to best support this model. The resulting Toluca-Cape Town Declaration (2024) affirmed diamond OA as a global priority for equitable scholarly communication. These summits follow from, and have been complemented by, the highly-visible efforts from the European context, including the large-scale study mapping the global diamond OA landscape (Bosman et al., 2021), an Action Plan setting out shared priorities around quality, sustainability, and capacity building (Ancion et al., 2022), dedicated funding programmes from national research funders including the Dutch Research Council (NWO, 2025) and the German Research Foundation (DFG, 2025), and the funding of diamond OA projects by the European Commission, including DIAMAS, CRAFT-OA, and ALMASI.

Despite all this recent attention and interest, diamond OA is in no way a new invention. In fact, the earliest OA journals on the web charged neither authors nor readers as a matter of course—not necessarily as a principled stance, but because no other OA model yet existed. A second model emerged only later, when APCs offered a way to provide free access without diminishing publisher revenues—and with it came the need to distinguish the two: “gold OA” came to encompass fee-based models and the no-fee model became known as “diamond” (Fuchs & Sandoval, 2013; Haschak, 2007). Latin America, in particular, embraced the diamond model through a combination of institutional support and regional initiatives, since journals started coming online en masse in the early 2000s (Alperin & Fischman, 2015; Alperin & Rozemblum, 2017). Diamond OA remains the dominant model there, not only among OA journals, but across all journals from the region. Yet despite this long history, global attention, and recent investments, the model is by no means secure.

Against what appears to be strong and growing momentum behind diamond OA, this paper documents a pattern of journals abandoning the model in favour of APC-based or subscription publishing. Our goal is to capture what might otherwise be dismissed as anecdotal setbacks to an otherwise encouraging trend. To assist us, we introduce the term 'diamond fracture' to describe the transition of a journal from a model that was free for readers and authors, to a subscription-based, hybrid, or APC-based gold OA model. By naming the phenomenon and bringing empirical evidence to bear, we aim to draw attention to a dynamic that current policy discussions have largely overlooked: that the challenge facing the diamond OA community is not only one of adoption, but also one of retention.

Drawing on multiple complementary data sources, we identify and characterize 440 fractures, examining how they are distributed across publisher types, disciplines, and regions; what models journals transition to; at what APC price points they enter the author-pays market; how old journals are when they fracture; and how publication volume shifts in the period surrounding the transition. That these fractures occur at all raises questions about the long-term stability of diamond OA as a publishing ecosystem. Journals that introduce APCs, even at low initial price points, enter a market with well-documented inflationary dynamics: author-pays charges have consistently risen above inflation in a market where prestige incentives insulate publishers from normal price competition (Haustein et al., 2024; Khoo, 2019). As APC-based publishing becomes further entrenched through funder mandates and commercial OA incentives, the conditions that push underfunded journals toward fracture are likely to grow more prevalent, not less. Treating each fracture as an individual failure obscures what is in fact a structural pattern, and one that calls for structural responses that make diamond OA viable over the long term and across a broader set of contexts.

# Previous research

## Identifying diamond OA journals

Establishing the scale and composition of diamond OA has proven methodologically challenging. Many bibliometric studies rely on journals listed as no-APC in the Directory of Open Access Journals (DOAJ) as a proxy for the global diamond OA journal population, but a substantial share of diamond OA journals have never applied for inclusion or do not meet its criteria (Laakso et al., 2025). The OA Diamond Journals Study (Bosman et al., 2021) addressed this directly, combining a purpose-built global inventory with an extensive survey of journal operations, and estimated between 17,000 and 29,000 active diamond OA journals worldwide—far more than indexed sources suggest. Where Bosman et al. (2021) capture the scale of the population at one point in time, Crawford's series of OA books—despite being limited to DOAJ-indexed journals—provides the most comprehensive longitudinal, article-level perspective on how diamond OA has evolved. He finds that diamond output has been broadly stagnant at around 400,000 articles per year between 2020 and 2024, while APC-based output grew from approximately one million to 1.4 million articles over the same period (Crawford, 2025). This divergence is striking given that diamond OA journals outnumber APC-based journals by nearly two to one (13,083 versus 7,026). A recent literature review by Taubert (Taubert, 2026), covering 66 publications from 2020 to 2025, finds that studies have made considerable methodological effort to compensate for limited database coverage often by combining data from multiple sources, but notes that studies would benefit from fewer a priori assumptions about the nature of diamond OA.

A recurring methodological challenge in diamond OA research is the practice among some commercial publishers of delaying APC introduction while newly launched journals are establishing themselves and seeking indexing, such that a journal may appear diamond OA at the moment of observation without any long-term commitment to the model (Khoo, 2019; Simard et al., 2024). Our definition of diamond fracture is designed to capture this pattern rather than exclude it.

## Changes in journal publishing models

Studies on publishing model changes across larger populations of journals are scarce, whether examining journals converting to OA or moving away from it. Studies of flips toward OA have focused predominantly on citation effects (Bautista-Puig et al., 2020; Li et al., 2018; Momeni et al., 2023). Research on the reverse direction is thinner still, and what exists has largely treated reverse flips as incidental findings rather than the primary object of study. Sotudeh and Horri (2007) found that 22 journals had transitioned from fully OA to a delayed, hybrid, or closed-access model—a byproduct of a broader study of OA journal evolution. Björk, Shen, and Laakso (2016) similarly encountered reverse flips in passing, finding 12 closures to subscription among a longitudinal cohort of 250 OA journals followed through 2002–2014. The first study to treat reverse flips as its central focus is Matthias, Jahn, and Laakso (2019b), who combined multiple data to identify 152 journals that had moved from OA to subscription-based or hybrid publishing. Across all three studies, a consistent methodological picture emerges: information about publishing model changes is difficult to trace. Crucially, none treats the departure from diamond OA specifically as a distinct phenomenon—which means the field currently lacks a conceptual and empirical foundation needed to understand how, where, and under what conditions diamond fractures occur. In this paper, we seek to address this limitation by tackling the following research questions specifically about what we term diamond fractures:

1) **What types of journals have undergone diamond fractures?**
    a) Among what types of publishers, disciplines, and geographical regions can journal fractures be found?
    b) At what point in their lifecycle do journals transition?
2) **What becomes of diamond journals that have fractured?**
    **a)** Do diamond fractures coincide with changes in ownership, and what types of publishers precipitate fractures?
    **b)** What publishing models do fractured journals transition to?

**c)** Where journals transition to APC-based models, at what price level do they enter the author-pays market? And how does this relate to the publishing context?

**d)** How does publication volume change following a fracture?

# Methods

## Data collection

To address these questions, we built a curated dataset capturing core information for each journal: ISSN, title, year founded, last year as diamond OA, year of fracture, publisher pre- and post-fracture, and APC and OA status post-fracture. Because no single source tracks OA status and APC information across time, we drew on five complementary data sources: the Publisher OA Portfolios dataset (Matthias, 2020), the ScholCommLab APC dataset (Matthias et al., forthcoming), an existing dataset of reverse-flip journals (Matthias et al., 2019a), DOAJ removal lists covering 2014–2026 (DOAJ, n.d.-a, n.d.-b), and a survey of journals using OJS. The sources varied considerably in completeness and scope, and each required its own extraction approach. The ScholCommLab dataset was the most complete, requiring primarily automated APC checks; others, like the DOAJ removal list, provided only basic journal metadata, meaning pre- and post-fracture information had to be gathered independently for each entry. Most candidate journals therefore underwent manual verification before inclusion in the final dataset.

## Identifying diamond fractures

Two of our sources tracked APC and OA status over time. The *Publisher OA Portfolios dataset* documents the OA journal portfolios of Elsevier, Sage, Springer Nature, Taylor & Francis, and Wiley spanning from each publisher's first fully OA journal through 2020, with additional data for select years from Cambridge University Press, Copernicus, Hindawi, and Oxford University Press—close to 70,000 entries in total. The *ScholCommLab APC dataset* tracks APC pricing from 2019–2025 across 14 major publishers, comprising nearly 70,000 entries. For both, we applied a consistent logic to detect fractures: for each journal, we located the most recent year in which it might have been diamond OA, defined as having an APC of zero or missing and a non-hybrid OA classification, and searched for a subsequent year with a positive or missing APC, or a switch to hybrid OA. We additionally flagged journals that disappeared from either dataset following their last potential diamond year, as this may indicate a transition to closed access or a publisher transfer. This yielded 746 candidates from the Publisher OA Portfolios dataset and 255 from the ScholCommLab data, with a further 89 identified by cross-checking journals last appearing as diamond OA in the Publisher OA Portfolios dataset against subsequent APC records in the ScholCommLab data.

The other three sources required different approaches. The *reverse-flip dataset* is an existing curated collection of 152 journals identified as having transitioned from OA to subscription-based or hybrid models; retaining only those with a confirmed APC of zero in the year prior to their transition yielded 73 candidates. The *DOAJ removal lists* document journals withdrawn from the Directory of Open Access Journals; the two lists consulted—covering March 2014 to January 2024 (5,280 journals) and February 2024 onwards (1,933 entries as of March 2026)—yielded 211 candidates after keeping only entries where the stated removal reason indicated a loss of OA status. Finally, a *survey of journals using OJS* identified 119 candidates through two questions: whether journals currently charge APCs, and whether they always have.

## Creating a unified dataset

Across all sources, we identified 1,404 candidate entries. We then identified and resolved duplicates, leaving 1,308 unique journals. Where the same journal appeared across multiple sources, we used the duplicate entries to cross-validate and complement each other, retaining the most complete information across records. For instance, where a DOAJ entry simply recorded a loss of OA status, a matching record in another dataset might provide the pre-fracture APC and post-fracture access model needed to confirm the fracture. We resolved conflicts, such as differing APC values or fracture years, manually against the Wayback Machine.

To confirm a fracture, we required evidence of two things: that the journal had operated as diamond OA at some point, and that its access model subsequently changed to subscription-based, hybrid OA, or gold OA. For 263 journals the source records already showed both—a diamond year (APC of zero, non-hybrid OA) followed by a year with a positive APC or a switch to hybrid—and needed no further checking. The remaining 1,045 had incomplete records, which we verified manually against archived journal websites in the Wayback Machine, recording pre- and post-fracture URLs as evidence. To count as diamond, a journal website had to state explicitly that it charged no fees; the mere absence of any mention of APCs was not sufficient. This confirmed 177 and discarded the rest.

The final dataset contains 440 confirmed fractures and includes the following fields for each journal: ISSN, journal title, last diamond year, pre- and post-fracture APC, pre- and post-fracture publisher, post-fracture OA model, fracture year. For all but 95 journals, we were able to confirm the pre- and post- fracture timeline within a year of each other (Figure 1). Of these, 55 fell within two years and 40 within three to six.

A dataset containing the 440 confirmed fractures and their characteristics is available at the ScholCommLab's Dataverse (Matthias et al., 2026).

**Figure 1**

*Precision of fracture dating*

**Figure 1. Precision of fracture dating**

Number of journals by gap between last diamond year and fracture year (n = 440).

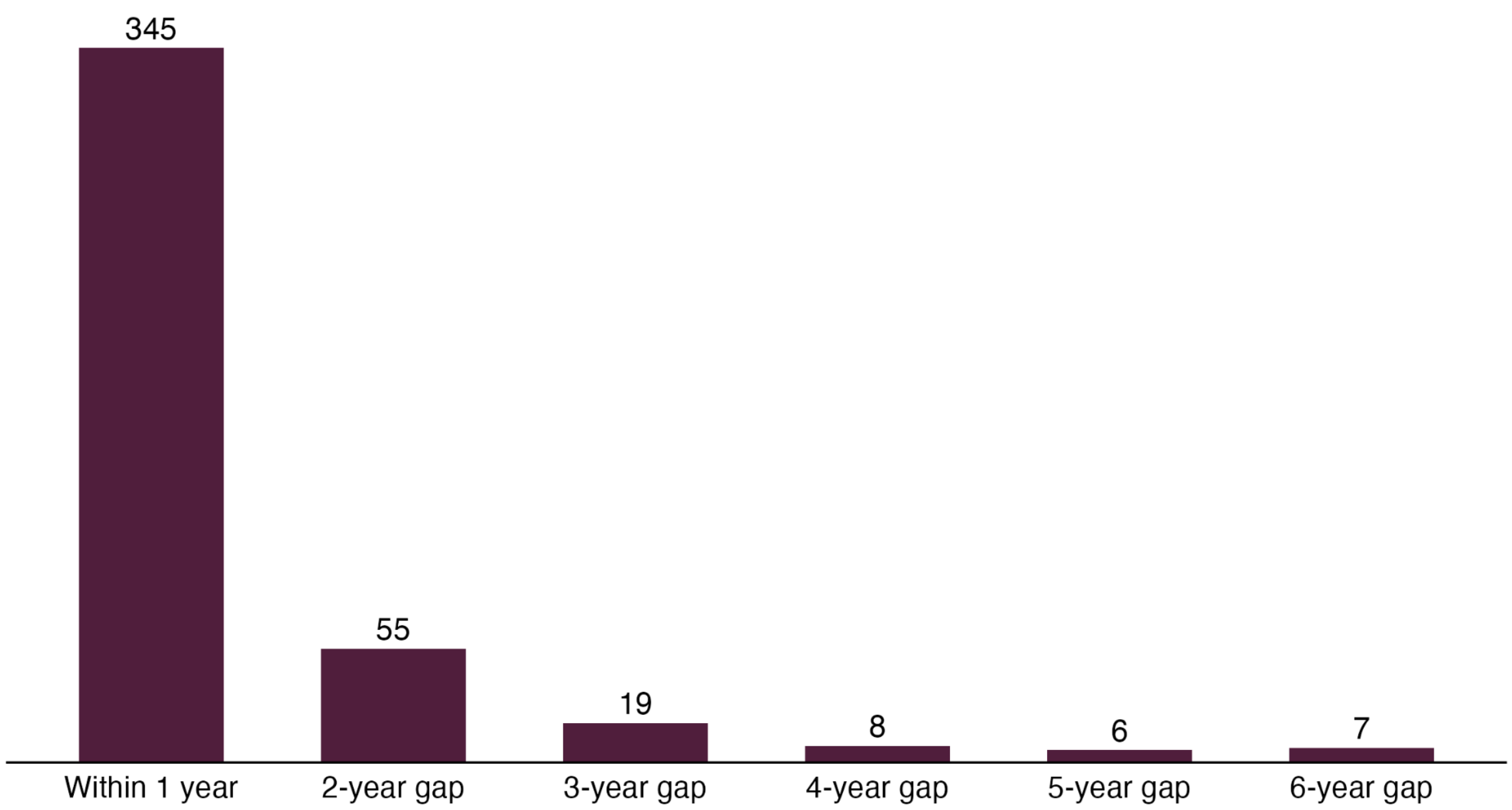

To aid in describing the fractures, we classified the publisher of each journal as either "professional," a "research organization," or a "scholarly society" as per Nazarovets, Laakso & Taşkın (2026). We used the primary domain field in OpenAlex to identify the primary discipline of each journal, and the ISSN Portal to determine its country, prioritizing the country of the issuing body over the listed country. Where the record was inconsistent or ambiguous—different countries across formats, a URL pointing elsewhere, or indication of a partnership—we verified against the journal website. We also added a founding year for each journal, drawn from Ulrichsweb, the ISSN Portal, or confirmed through manual searches where it was not listed.

# Results

## What types of journals have undergone diamond fractures?

### Among what types of publishers, disciplines, and geographical regions can journal fractures be found?

Of the 440 confirmed diamond fractures, the distribution of publisher types, disciplines and regions reveals that the phenomenon is not specific to any one group (Figure 2). By far, professional publishing organizations were the most common pre-fracture type (n = 342), followed by research organizations (n = 67) and half as many society publishers (n = 31). The dominance of professional publishers may be indicative of a commercial orientation, or a greater reluctance to abandon diamond by scholar-led publishers.

Across disciplines, there is a more even distribution. Fractures were most prevalent in physical sciences (n = 151) and health sciences (n = 114), though not by a large margin. Notably, we identified fewer instances of journals from the life sciences (n = 72) than from the social sciences (n = 95), despite the significantly higher prevalence of APCs in that field. The lack of correspondence between the use of APCs in a given field and the number of fractures identified suggests that the phenomenon is subject to other systemic pressures.

What is most striking about the demographic of the journals that have undergone fractures is their geographic composition, with journals originating in Asia (n = 207) making up nearly half of the dataset, and European journals (n = 159) over one third. The concentration in Asia is particularly notable given that region's relatively recent explosion in publishing, and suggests that this expansion may have started out as diamond, but is quickly transitioning to other models. As with disciplines, there is little correlation between the prevalence of diamond publishing in a region and the number of fractures identified; Latin America—a region that publishes almost exclusively diamond OA—has seen relatively few fractures.

**Figure 2**

*Types of journals that have undergone diamond fractures*

**Figure 2. Types of journals that have undergone diamond fractures**
n = 440 confirmed fractures

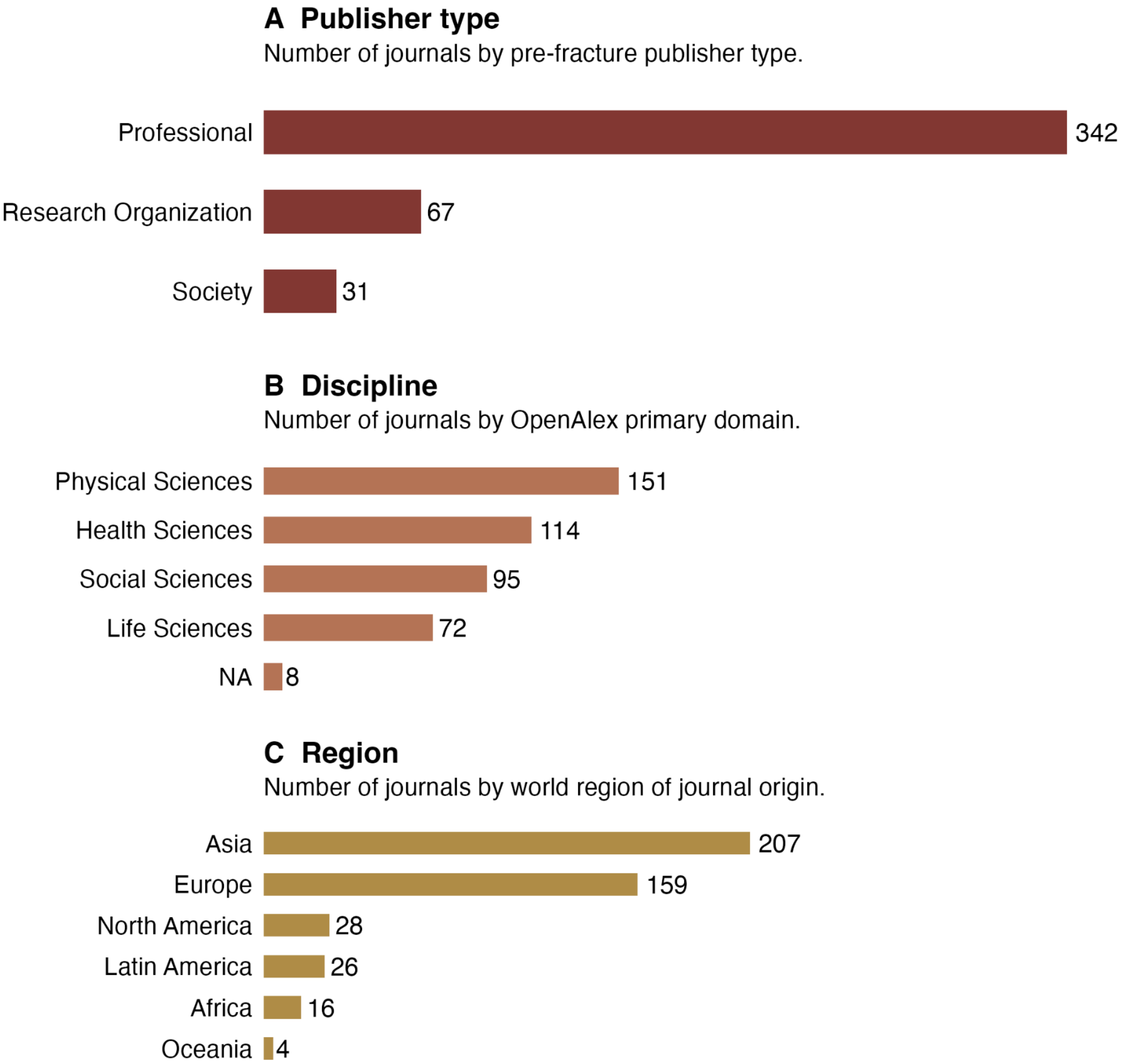

## At what point in their lifecycle do journals transition?

Diamond fractures are not a new phenomenon. We were able to uncover at least one fracture every year since 2009, although their occurrence has been accelerating significantly starting in 2017 and especially since 2020 (Figure 2A).

While the true rate of fractures for the earlier years may be higher than the data suggest, given the recency bias in the data sources we used to construct the dataset, this large increase in fractures in 5-8 years is unlikely to be an artifact of the data collection approach. This recent acceleration coincides with both the rapid expansion of the APC model, and the increasing policy and advocacy attention to diamond OA.

Perhaps surprising is that over half the observed fractures (n = 249) are concentrated among journals that had been in operation for at least nine years. In contrast, less than half that many journals (n = 133) were only recently established (1-5 years prior to the fracture). These data suggest that diamond fractures may not be a mechanism for those struggling to sustain themselves, but rather a systemic temptation to profit from well-established brands.

**Figure 3**

*Journal lifecycle at point of fracture*

**Figure 3. Journal lifecycle at point of fracture**

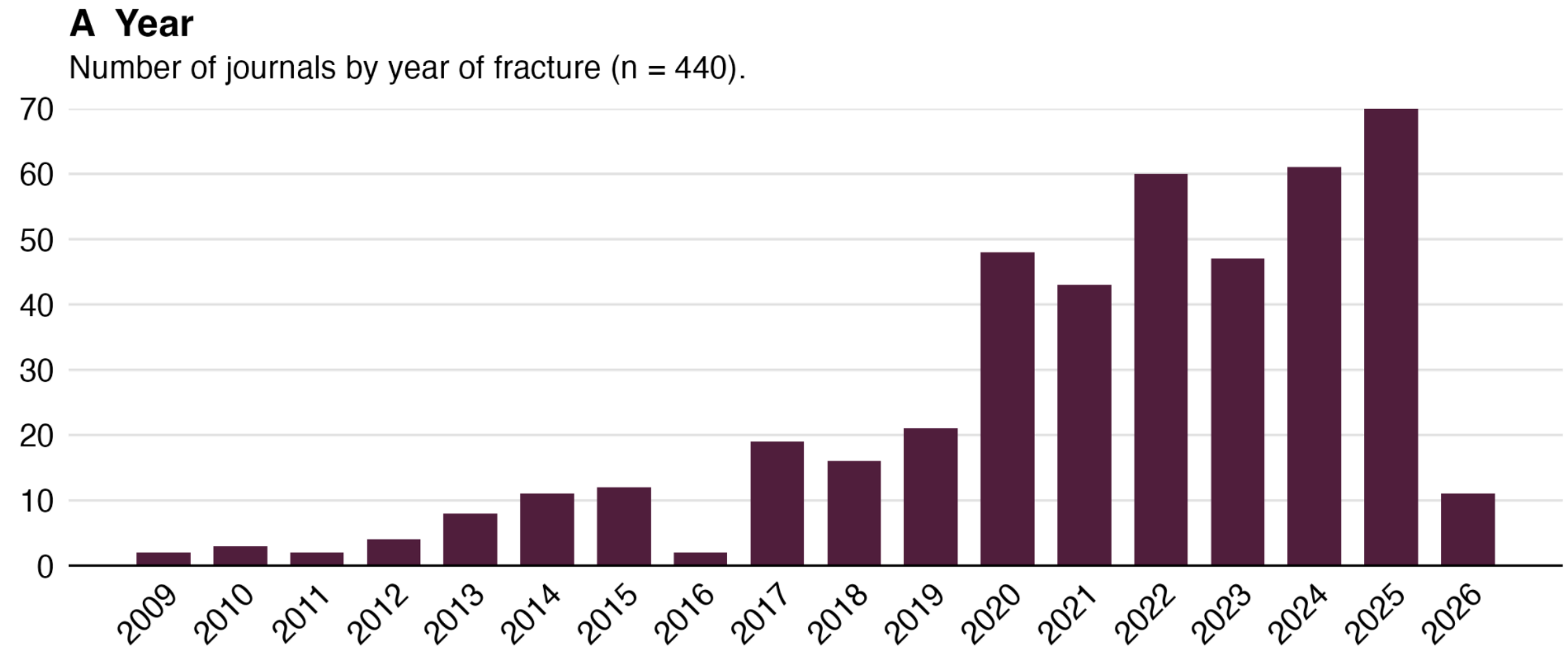

**B Journal age**

Number of journals by age at fracture, calculated as year of fracture minus founding year (n = 439; one journal excluded due to missing founding year).

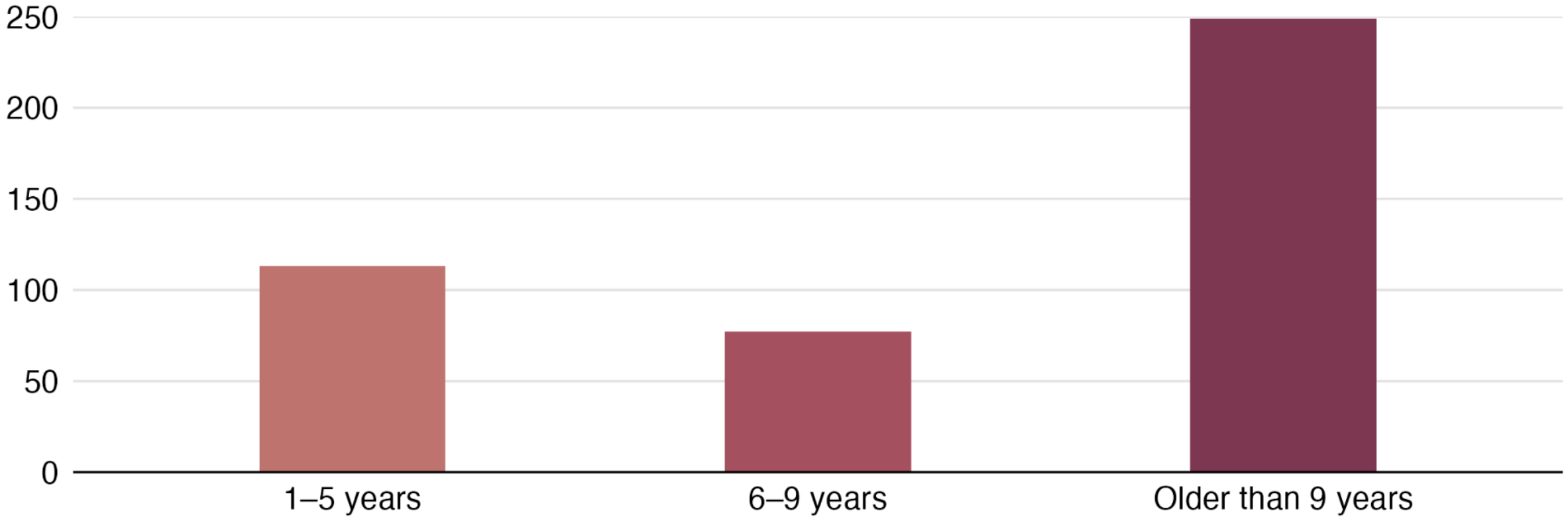

## What becomes of Diamond journals that have fractured?

### Do diamond fractures coincide with changes in ownership, and what types of publishers precipitate fractures?

Most diamond fractures occur without any change in ownership: 389 of 440 journals fractured while staying with the same publisher that had operated them as diamond OA (Figure 4). Of the 51 fractures that did coincide with an ownership change, 44 involved transfer to a professional publisher, whereas societies and research organizations accounted for just seven. Taken together, these patterns suggest that diamond fracture is predominantly an internal decision by publishers who chose to orient the journals toward revenue-generating models, and that when ownership does change, it consolidates further within the professional publishing sector.

**Figure 4**

*Changes in publisher ownership at fracture*

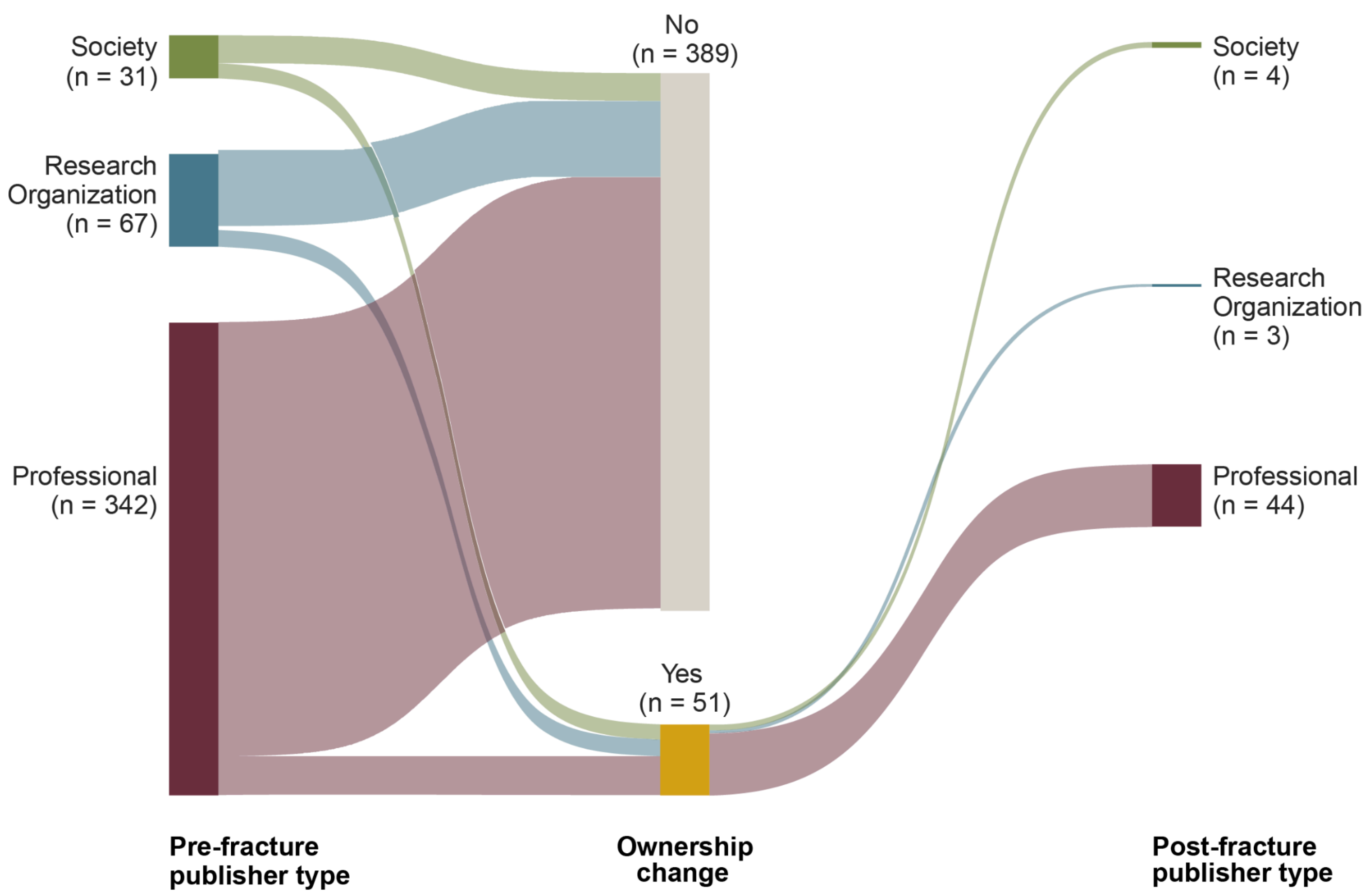

## What publishing models do fractured journals transition to?

The overwhelming majority of fractured journals transition to APC-based gold OA (n = 369), making it by far the most common destination. A further 52 adopted a hybrid model, and only 19 closed to readers entirely. This finding suggests that diamond fractures are not primarily a story of journals retreating behind paywalls, but a story of journals entering the author-pays market.

**Figure 5**

*Post-fracture publishing model*

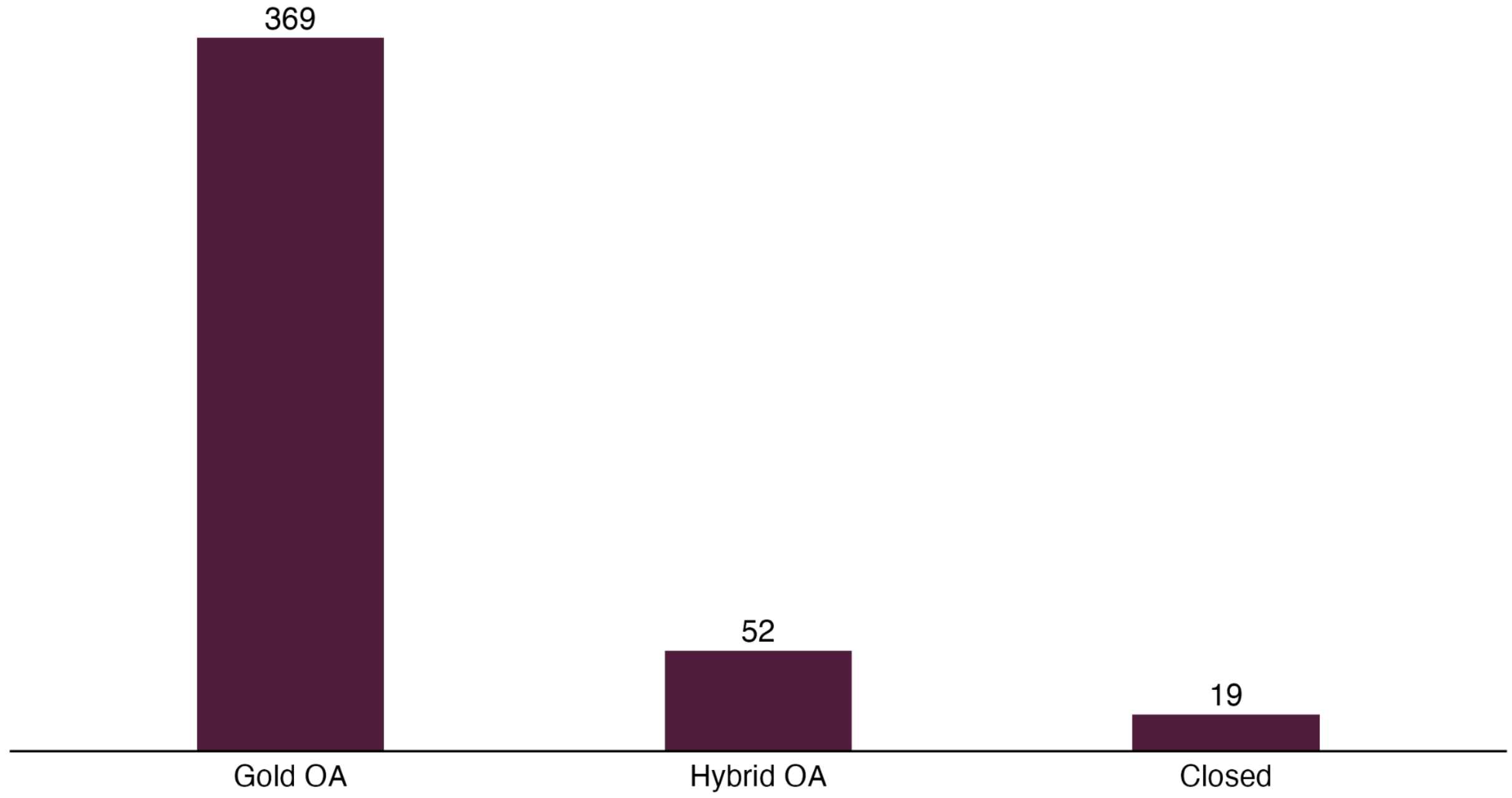

**Figure 5. Post-fracture publishing model**
Number of journals by publishing model after fracture (n = 440).

## Where journals transition to APC-based models, at what price level do they enter the author-pays market? and how does this relate to the publishing context?

Entry APCs among fractured journals range from $8 to $5,300, with the highest charges at *Energy & Environmental Materials* ($5,300) and *Cell Reports* ($5,000). Among the 407 journals with a known flat-rate APC (excluding 14 with page charges or unknown fees), entry price points vary substantially by publishing context. For journals transitioning to gold OA (n = 360), the largest concentration falls in the $1,001–2,000 range (36%), though prices are spread across the full distribution, with 14% entering below $100 (Figure 6A). Hybrid OA journals enter at markedly higher levels, with 70% charging $2,001–3,000 at entry. The sharpest contrast is by publisher type: professional publishers cluster in the $1,001–3,000 range (70% combined), while research organizations (65%) and societies (52%) enter predominantly below $100 (Figure 6B). Journals that fracture are not all entering the same market—professional publishers are pricing at levels competitive with established commercial OA, while non-commercial publishers are introducing token charges that may reflect a different logic, whether cost recovery or external pressure, rather than a straightforward commercial pivot.

**Figure 6**

*APC level at entry to author-pays market*

**Figure 6. APC level at entry to author-pays market**

n = 407 journals with a known flat-rate APC. Journals with page charges and unknown APCs excluded (n = 14).

**A Entry APC level by OA publishing model**

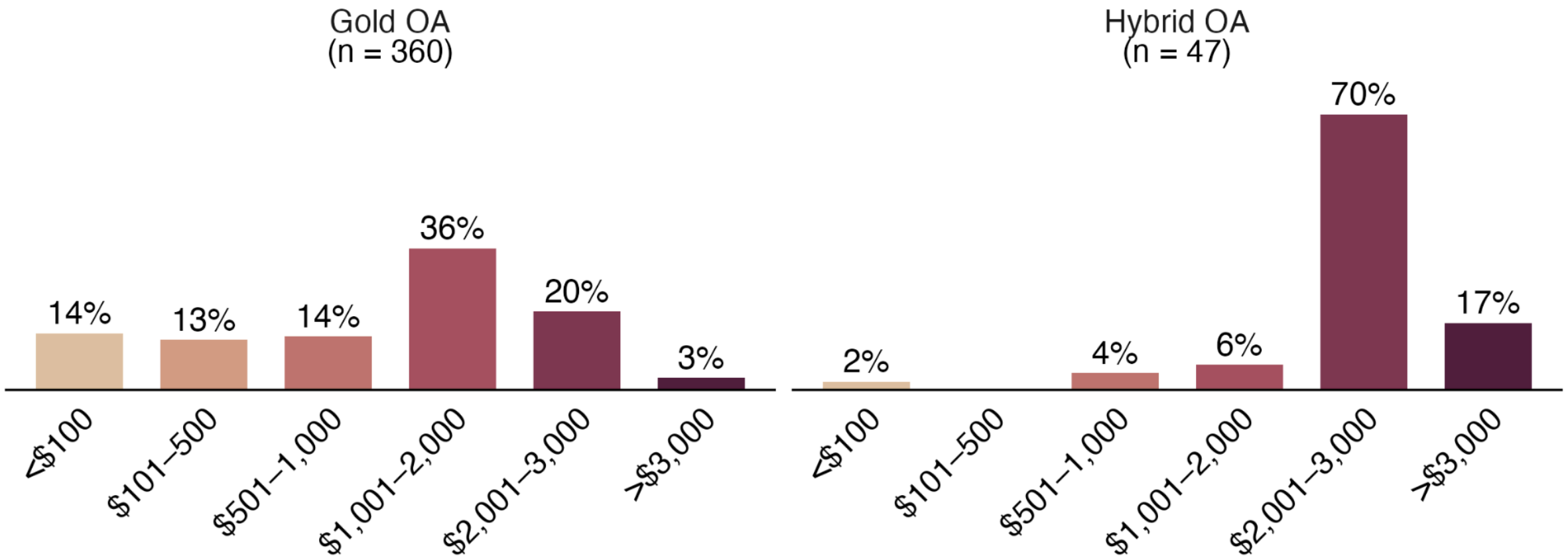

**B Entry APC level by publisher type**

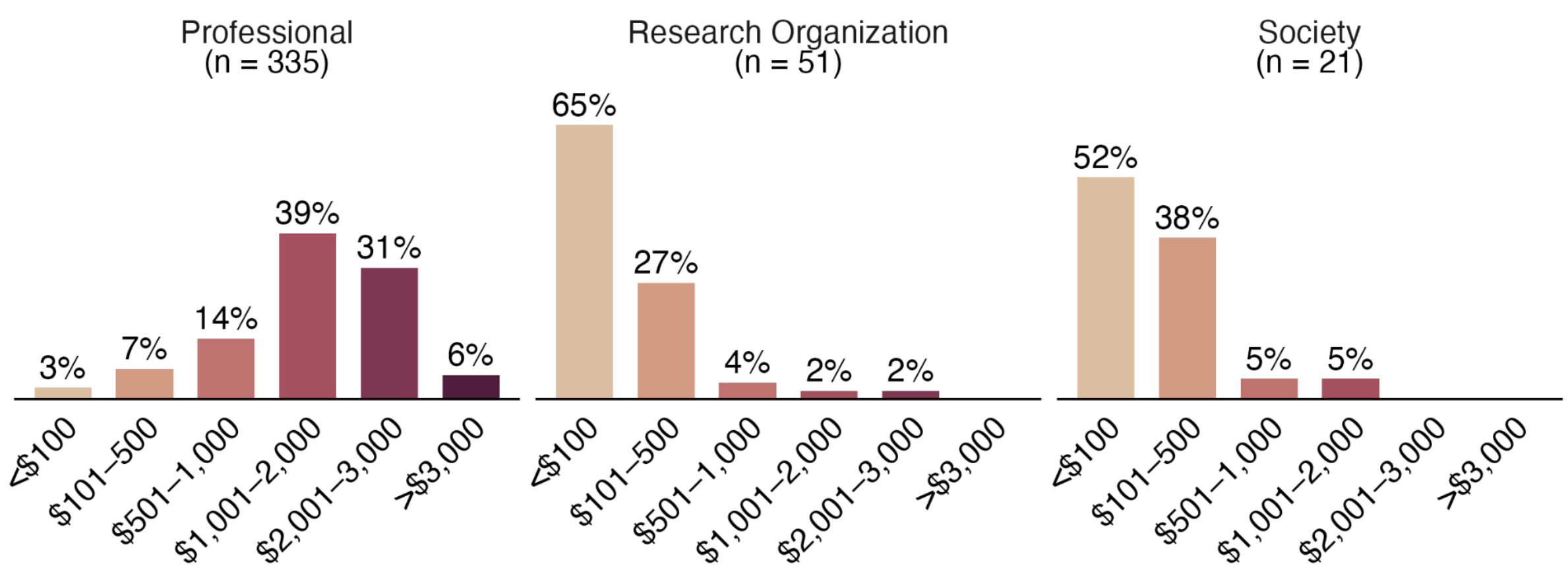

## How does publication volume change following a fracture?

Publication volume changes little following a fracture, with the median journal publishing roughly the same number of articles in the two years after fracturing as it did in the fracture year itself (Figure 7A). What changes is the variability: the distribution widens substantially post-fracture, indicating that while some journals grow, others contract sharply. The fracture year itself tends to be a moment of relative publishing strength—across the 302 journals with publication data across all time points, the median journal published roughly 16% fewer articles three years before fracturing than it did in the fracture year, with volume climbing steadily toward that peak. What stands out is less this modest growth than the scale at which these journals were already operating. Figure 7B shows their average output in the years leading up to fracture: about half published fewer than 50 articles per year, but a substantial share sat well above the diamond OA norm of 34 articles annually (mean) reported by Bosman et al. (2021), and two exceeded 500. Fractures, in other words, are concentrated among journals publishing at a scale that sits outside the typical diamond range—a pattern that echoes Matthias, Jahn, and Laakso's (2019b) finding that reverse-flipped journals had above-average output before flipping that remained stable afterward. At that scale, charging APCs may become either necessary to sustain operations or too financially attractive to forgo.

**Figure 7**

*Publication volume*

**Figure 7. Publication volume**

**A Publication volume relative to fracture year**

% difference from fracture year (= 0%). Median with IQR ribbon. n = 302 journals with complete data across all time points.

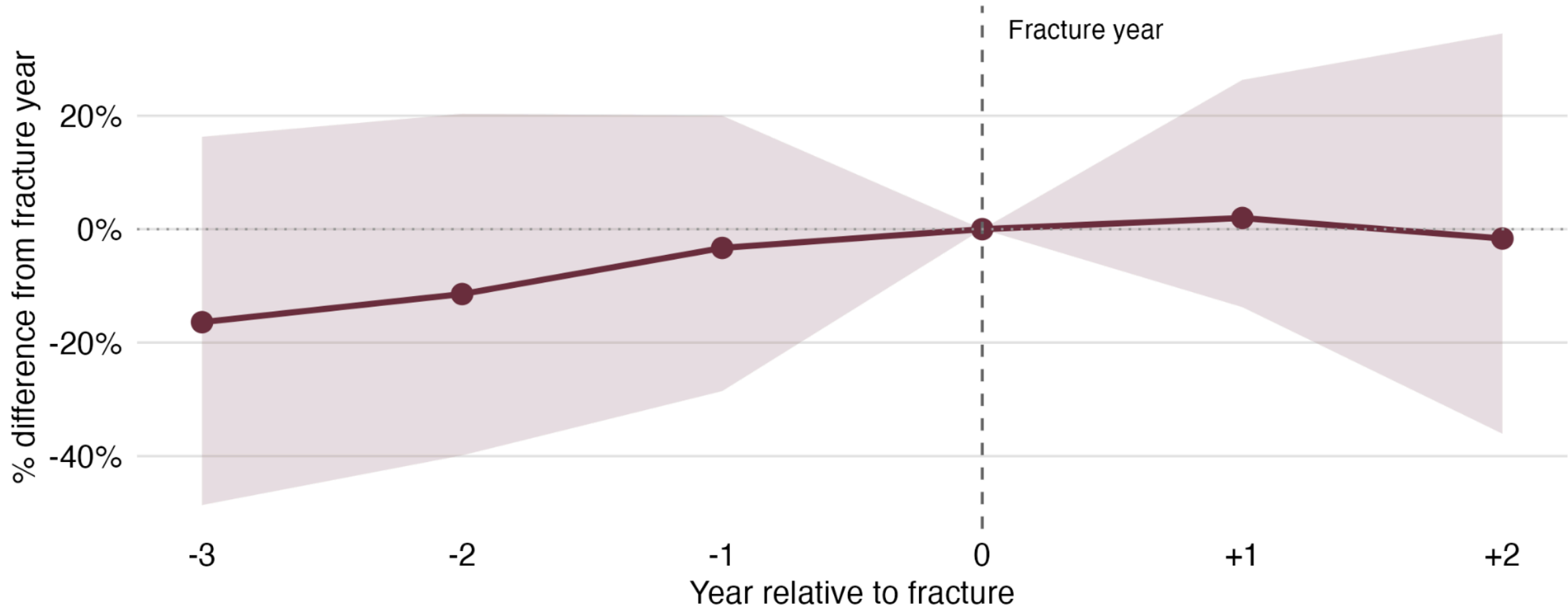

**B Publication volume before and at fracture year (grouped)**

Number of journals per article count bucket (n = 302). Faded bars = pre-fracture average (years –3 to –1). Solid bars = fracture year (year 0).

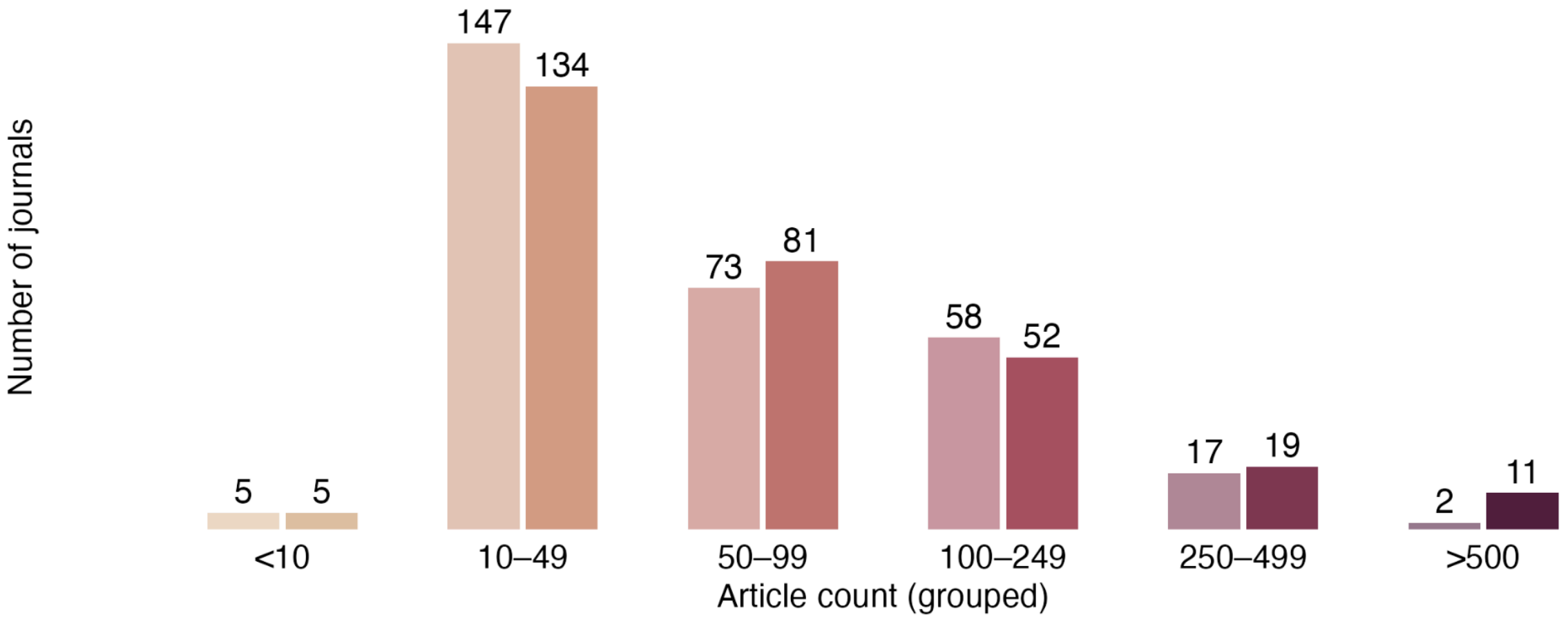

# Discussion

Despite all of the interest in diamond OA as a desirable alternative to APCs, how, when, where, and why journals abandon the model remains largely unexamined. In naming the phenomenon and documenting 440 cases spanning 2009 to 2026, this study offers the conceptual and empirical foundation necessary to understand and critically examine this aspect of OA publishing. Just as research on reverse flips brought awareness about journals leaving OA altogether (Matthias et al., 2019b), this study brings the same attention to journals leaving diamond OA and highlights that the challenge facing diamond OA requires more than attracting new journals to the model—it also requires protecting existing ones. This two-sided challenge calls for a different policy response and further institutional commitments to the model.

The fractures we document are not the complete set, nor are they necessarily representative of every journal that has fractured. Identifying diamond journals is a standing challenge in itself: as Taubert (2026) notes, no single source comprehensively covers the current and past population, and studies routinely combine sources to compensate. Detecting the journals that leave the model inherits this difficulty. Our sources are biased towards large publishers, DOAJ-indexed journals, and journals running on OJS, and their coverage has a strong recency bias; hence it is likely we will have missed smaller, independent, and older fractures. Nevertheless, the findings both support and challenge common assumptions about the dynamics surrounding diamond OA publishing.

Diamond fractures are not new—we identified at least one in every year since 2009—but they do seem to be accelerating, with the majority occurring from 2020 onward and 2025 marking the highest annual count in the dataset. Nearly all fractured journals (n = 369, 84%) transitioned to APC-based gold OA, with a further 52 adopting hybrid models and only 19 closing to readers entirely. Fracturing, then, rarely means fully retreating behind a paywall; it means entering the growing author-pays market (Haustein et al., 2024). As APCs have become an industry norm, charging authors has become an economically tempting and viable path even for journals that had never charged a fee. We do see two distinct strategies for APC pricing, perhaps indicative of different motivations: professional publishers enter the market at prices competitive with established commercial APC rates ($1,001–3,000; 70%), as though their journals had always charged, while research organizations (65%) and societies (52%) introduce charges below $100. However, given the well-documented inflationary dynamics of APCs (Haustein et al., 2024; Khoo, 2019), even modest APCs should be regarded with caution. A low APC today is unlikely to remain low in the long run, especially if the journal in question is successful.

Moreover, we found that fractures span all disciplines, regions, and publisher types, but concentrate where commercial orientation is strongest: professional publishers account for 78% (n = 342). Most fractures are also internal processes—88% occurred without any change of publisher, indicating leaving the diamond model is a business choice rather than a shift in publishing responsibilities. When ownership did change, it consolidated the pattern further, with most transfers moving journals to professional publishers.

The relative scarcity of fractures in Latin America—despite diamond OA's dominance in the region—further points to a relationship between fractures and commercial publishing. It appears that the presence of regional infrastructure (e.g., Redalyc and SciELO) and high involvement from universities continue to be enough to stave off fractures—at least for existing journals. The fragility of the ecosystem is making itself known elsewhere, with one study showing that the number of Latin American journals charging APCs doubled between 2021 and 2023, from 123 to 249 (Córdoba González, 2024). That is, it appears that structural supports that have protected the region from APCs are holding, but it is reasonable to conclude that, without structural support, the path of least resistance runs toward commercialization.

The shifts away from diamond were not accompanied by the marked increase in publishing volume, as might be expected from a pure profit motive. The financial incentives embedded in profit-driven publishing would lead us to expect otherwise, as would cases of mass editorial board resignations over pressure to increase output in order to maximize revenue (van Walsum et al., 2026). Yet across the population of fractured journals we found no such pattern: the median journal published roughly as many articles in the two years after fracturing as in the fracture year itself. Some journals did grow sharply, but they were offset by others that contracted, so the distribution widened without the median shifting. Either volume growth does not follow fracture, or it follows at a lag of two years or more. Crawford's (2025) longitudinal analysis of DOAJ journals found diamond article output flat between 2020 and 2024 while APC-supported output grew—a divergence that could reflect author migration or differential growth among journals, but to which our findings suggest fractures contribute as well.

The data suggest at least two distinct mechanisms leading to fractures. For relative newcomers (journals operating for less than five years prior to the fracture; n = 133), we see the pattern of publishers forgoing APC revenue as a promotional strategy, while a new journal builds indexing and reputation, then introduce them once the journal is established (Khoo, 2019; Simard et al., 2024). However, over half of journals in our data (57%, n = 249) fractured after at least nine years of operation. These are not newly founded titles using the absence of APCs to attract manuscripts while still establishing themselves—they are journals whose reputational capital and established brand has become commercially exploitable. The temptation and relative ease of adopting APCs is particularly evident for these journals, given that such established journals would be strong candidates for collective funding models, such as Subscribe to Open.

This research and its associated dataset present several directions for further work. The first is to examine the circumstances and motivations behind fractures by interviewing the editors and publishers involved. A second direction concerns the journals that close rather than fracture. Leaving the diamond model is one way a diamond journal can end; ceasing to publish is another, and this study measured only the first. How often diamond journals shut down, and how that compares to the rate of fracturing, remains open. The third follows from the split in publishing volume. Our population-level analysis cannot explain why some fractured journals expanded while others contracted—whether through shifts in submissions, editorial practice, or other factors—a question that would require a more detailed investigation.

Our research adds a further dimension to the complex and changing landscape of scholarly publishing. As momentum behind diamond OA builds alongside calls to lower or eliminate APCs (Cancer Research UK, 2026; National Institutes of Health, 2025; Office of Management and Budget, 2026) and rising concerns over Read-and-Publish agreements, our findings underscore the need to preserve the diamond publishing that already exists. Doing so requires more than opposition to APCs—it requires supporting underlying infrastructures and investing in the ecosystem itself.

Diamond fractures are not failures of individual journals. They are the predictable outcome of a publishing ecosystem that rewards commercialization and provides insufficient structural alternatives. Understanding them as such is the first step toward addressing them.

## Data availability

The data that support the findings of this study are openly available in Harvard Dataverse at https://doi.org/10.7910/DVN/6P2C7X.